\documentclass[twocolumn,10pt]{article}
\usepackage{sidecap,graphicx}
\title{Predictions of Quark-Gluon String Model for pp at LHC}
\author{A.B. Kaidalov$^1$ and M.G. Poghosyan$^2$\\
$^1$Institute of Theoretical and Experimental Physics, 117259 Moscow, Russia\\
$^2$Universit\'a  di Torino/INFN, 10125 Torino, Italy
}                     
\date{}
\begin{document}
\maketitle
\begin{abstract}
Soft multiparticle production processes in hadronic collisions 
are considered in the framework of the Quark-Gluon Strings Model 
and the model predictions are compared with data from S$p\bar{p}$S 
and Tevatron. Predictions for LHC energies are given.
\end{abstract}
\section{Introduction}
\label{intro}
In hadronic collisions soft interactions are dominant. 
These are characterized by large distances ($r\sim 1/\Lambda_{QCD}$) 
and the pQCD cannot be used for their study. 

Regge-pole theory is the main method for description of high-energy 
soft processes and the Pomeron is the main object of this approach, 
which is associated with cylinder-type diagrams of $1/N$ expansion 
in QCD. $1/N$ expansion is a dynamical expansion and the speed of 
convergence depends on the kinematical region of studying process. 
At very high energies many terms of the expansion (multi-pomeron 
exchanges) should be taken into account.

The models of Quark-Gluon Strings  (QGSM) \cite{ref1} and Dual-Parton 
Model (DPM) \cite{ref2}, are based on nonperturbative notions, combining 
$1/N$ expansion in QCD with Regge theory and using parton structure of 
hadrons. Here we concentrate on QGSM, which has been successfully applied 
to many different problems of strong interactions:  hadronic mass spectrum 
\cite{ref3}, widths of resonances \cite{ref4}, relations between the total 
cross-sections, residues of Regge poles \cite{ref4}, behavior of hadronic 
form factors \cite{ref5}, baryon number transfer 
\cite{ref6}, and multiparticle production at high energies \cite{ref1,ref6,ref7,ref8}.

In this paper we briefly describe QGSM, compare its predictions with 
S$p\bar{p}$S and Tevatron data on charged particles pseudorapidity and 
multiplicity distributions and give predictions for LHC.

\section{QGSM}
\label{sec:1}
In QGSM the production of a particle is defined through production of showers 
and each shower corresponds to the cut-pomeron pole contribution in the elastic 
scattering amplitude. For hadron-hadron collisions the cross-section of $n$ 
cut-pomereon exchange (among arbitrary number of uncut pomerons) is calculated in 
``quasi-eikonal'' approximation and has the following form \cite{ref9}:
\begin{equation}
\label{Eq:Eq1}
\sigma_{n}(\xi)=4\pi\frac{\lambda}{nC}\left[ 1-exp\{ -z\} \sum_{l=0}^{n-1}\frac{z^{l}}{l!}\right].
\end{equation}
Where $z=C\gamma/\lambda\exp\{\Delta\xi\}$, $\lambda=R^2+\alpha_P^{\prime}\xi$, and $\xi=\ln(s/s_0)$. 
The values of parameters $\gamma$, $\lambda$, $R^2$ and $\alpha_P^{\prime}$ which characterize the 
residue and the trajectory of the pomeron are found from fit to data on $pp$ and $p\bar{p}$ total 
interaction and elastic scattering cross-section in Ref. \cite{ref10}: $\gamma$ = 2.14 GeV$^{-2}$, 
$R^2$ = 3.3 GeV$^{-2}$, $\Delta$ = 0.12, $\alpha_P^{\prime}$ = 0.22 GeV$^{-2}$. The parameter $C$, 
which is related to small-mass diffraction dissociation of incoming hadrons during the rescattering 
is equal to 1.5 \cite{ref11}. The sum of $\sigma_n$ ($n$ = 1,2,3, $\dots$) defines the cross-section 
of non-diffractive interaction.

The association of the pomeron with cylinder-type diagrams leads to the fact that in a single cut-pomeron 
diagram there are two chains of particles (strings). Analogously, in case of $n$ cut-pomerons there are 
2$n$ chains. Inclusive distribution in rapidities and multiplicity distribution of particles for 
non-diffractive events can be expressed in terms of rapidity distributions, $f_n(\xi,y)$, and multiplicity 
distribution, $W_n(\xi,N)$, for 2$n$-chains \cite{ref1}:
\begin{eqnarray}
\label{Eq:Eq2}
\frac{d \sigma(\xi)}{dy }= \sum_{n} \sigma_{n}(\xi) f_{n}(\xi,y),\\ 
\sigma(\xi, N) = \sum_{n}\sigma_n(\xi) W_n(\xi, N). \nonumber
\end{eqnarray}
In order to calculate the distribution of hadrons produced during the fragmentation of the 
strings one should take into account that the hadron can be produced in each of the 2$n$-chains 
and the (di-)quarks , which stretch the strings, carry only a fraction of energy of the incoming 
protons. In these terms, the inclusive spectra can be written as convolutions of the probabilities 
to find a string with certain rapidity length and the fragmentation functions, which defines the 
distribution of hadrons in the string breaking process. For $pp$ collisions the function $f_n(\xi,y)$ 
is written as follows \cite{ref1}:
\begin{eqnarray}
\label{Eq:Eq3}
f_n^h(\xi,y)=a^h\left[F^{h(n)}_{q_{val}}(x_+)F^{h(n)}_{qq}(x_-)
+F^{h(n)}_{qq}(x_+)F^{h(n)}_{q_{val}}(x_-) \right.\nonumber \\
\left.+2(n-1)F^{h(n)}_{q_{sea}}(x_+) F^{h(n)}_{\bar{q}_{sea}}(x_-)\right],
\end{eqnarray}
where $x_{\pm} = (1/2)[\sqrt{x_T^2+x^2} \pm x]$, $x_T=2m_T^h/\sqrt{s}$ and $x$ is the
Feynman-$x$ of produced hadron $h$. $a^h$ is the
density of hadrons $h$ produced at mid-rapidity in a single chain and its value is
determined from experimental data.  In fact these are the only free parameters in
the model that are fixed from fit to data. In articles \cite{ref1},\cite{ref6} and 
\cite{ref7} the following values are found for them: $a^{\pi}$ = 0.44, $a^K$ = 0.055, 
$a^{p}$ = 0.07.
The first two terms in Eq.~(\ref{Eq:Eq3}) correspond to the chains, which connect 
``valence''quarks and di-quarks\footnote[1]{In case of $p\bar{p}$ collisions they 
must be replaced by 
$F_{q_{val}}^{h(n)}(x_+)F_{\bar{q}_{val}}^{h(n)}(x_-)+F_{qq}^{h(n)}(x_+)F_{\bar{q}\bar{q}}^{h(n)}(x_-)$. 
This change is in order to have white objects required by QCD.}, and the last term corresponds to chains connected 
to the sea quarks-antiquarks. 
Here we do not consider the contribution of sea di-quarks and anti-diquarks, their 
contribution is important in the spectra of anti-baryons produced in the forward region 
\cite{ref12}. The functions $F_i^{h(n)}(x)$ being a convolution of the structure $\psi(x)$, 
and fragmentation $G(x)$ functions are writen as follows:
\begin{eqnarray}
F_{q_{val}}^{h(n)}(x_{\pm})=
\frac{2}{3}\int_{x_{\pm}}^{1}dx\psi_{u_{val}}(n,x)G_{u}^{h}\left(\frac{x_{\pm}}{x}\right) \nonumber\\
+\frac{1}{3}\int_{x_{\pm}}^{1}dx\psi_{d_{val}}(n,x)G_{d}^{h}\left(\frac{x_{\pm}}{x}\right),\nonumber\\
F_{qq}^{h(n)}(x_{\pm}) =
\frac{2}{3}\int_{x_{\pm}}^{1}dx\psi_{ud}(n,x)G_{ud}^{h}\left(\frac{x_{\pm}}{x}\right) \nonumber \\
+\frac{1}{3}\int_{x_{\pm}}^{1}dx\psi_{uu}(n,x)G_{uu}^{h}\left(\frac{x_{\pm}}{x}\right),\nonumber\\
F_{q_{sea}}^{h(n)}(x_{\pm}) =\frac{1}{2+\delta}\left[
\int_{x_{\pm}}^{1}dx\psi_{u_{sea}}(n,x)G_{u}^{h}\left(\frac{x_{\pm}}{x}\right)\right.\nonumber\\
+\int_{x_{\pm}}^{1}dx\psi_{d_{sea}}(n,x)G_{d}^{h}\left(\frac{x_{\pm}}{x}\right)+\nonumber\\
\left.\delta \int_{x_{\pm}}^{1}dx\psi_{s_{sea}}(n,x)G_{s}^{h}\left(\frac{x_{\pm}}{x}\right)\right].\nonumber
\end{eqnarray}
Where $\delta$ is the strangeness suppression parameter ($\delta \approx 1/3$). 

In the model, the structure and fragmentation functions are determined by the corresponding 
Regge asymptotic behaviors in the regions $x \rightarrow 0$ and $x \rightarrow 1$ and for the 
full range of $x$ an interpolation is done. 

The structure functions of a proton are parameterized as follows \cite{ref6,ref7}:
\begin{eqnarray}
\psi_{d_v}(x,n) = \psi_{d_s}(x,n) = C_{d}x^{-\alpha_{R}}(1-x)^{\alpha_{R}-2\alpha_B+n},     \nonumber\\
\psi_{u_v}(x,n) = \psi_{d_s}(x,n) = C_{u}x^{-\alpha_{R}}(1-x)^{\alpha_{R}-2\alpha_B+n-1},   \nonumber\\
\psi_{ud}(x,n) = C_{ud}x^{\alpha_{R}-2\alpha_B}(1-x)^{\alpha_{R}+n-1},  \nonumber \\
\psi_{uu}(x,n) = C_{uu}x^{\alpha_{R}-2\alpha_B+1}(1-x)^{\alpha_{R}+n-1},  \nonumber\\
\psi_{s}(x,n) = C_{s}x^{-\alpha_{R}}(1-x)^{2(\alpha_{R}-\alpha_B)-\alpha_{\phi}+n-1}.\nonumber
\end{eqnarray}
Where $\alpha_{R}$=0.5, $\alpha_B$=-0.5. $C_i$ are determined from normalization condition:
\begin{eqnarray}
\int_0^1 \psi_{i}(x,n) dx =1.\nonumber
\end{eqnarray}
General technique of constructing fragmentation functions is presented in Ref. \cite{ref13}. 
For instance, for fragmenting to a pion the parameterizations are (see \cite{ref1} and 
\cite{ref13}):
\begin{eqnarray}
G_{d}^{\pi^{+}}(z) = G_{u}^{\pi^{-}}(z) = (1-z)^{2-3\alpha_{R}+\lambda},\nonumber\\
G_{u}^{\pi^{+}}(z) = G_{d}^{\pi^{-}}(z) = (1-z)^{ -\alpha_{R}+\lambda},\nonumber\\
G_{s}^{\pi^{+}}(z) = G_{s}^{\pi^{-}}(z) = (1-z)^{1 -\alpha_{R}+\lambda},\nonumber\\
G_{ud}^{\pi^{+}}(z)= G_{ud}^{\pi^{-}}(z) = (1-z)^{\alpha_{R}+\lambda-2\alpha_R}(1-z+z^{2}/2),\nonumber\\
G_{uu}^{\pi^{+}}(z) = (1-z)^{\alpha_{R}+\lambda-2\alpha_B},\nonumber\\
G_{uu}^{\pi^{-}}(z) = (1-z)^{\alpha_{R}+\lambda-2\alpha_B+1}.\nonumber
\end{eqnarray}
We do not list fragmentation functions for kaons and (anti-)protons, they are taken from Ref. 
\cite{ref6,ref7}.

In hadronic interactions diffractive processes play an important role. Densities of 
produced particles in these processes are small, especially in the central rapidity 
region, but their cross-section is not negligible with respect to non-diffractive 
cross-section and they must be taken into account in calculations of characteristics 
of secondary particles. In Ref. \cite{ref14} single- and double- diffractive processes 
are described in terms of dressed triple-reggeon and loop diagrams. For instance, for 
calculating the rapidity distribution of particles produced in soft single-diffractive 
events the first equation in (\ref{Eq:Eq2}) must be replaced by (see \cite{ref8}):
\begin{equation}
\label{Eq:Eq4}
\frac{d\sigma}{dy}=\sum_{n}\int d\zeta \frac{d\sigma_n}{d\zeta}f_n(\zeta,y).
\end{equation}
Here $d\sigma/d\zeta$ is the derivative of single-diffraction dissociation cross-section, 
$\zeta=M^2/s$, and $M$ is the mass of the diffracted system. 

The model does not contain ``odderon''-type singularities with negative signature, which 
could lead to a difference of $pp$ and $p\bar{p}$  scattering, so at energies 
$\sqrt{s} \geq$ 100 GeV all characteristics of $pp$ and $p\bar{p}$  interactions coincide. 

\section{Numerical results}
QGSM has been used successfully to describe data on secondary hadron inclusive cross-sections 
integrated over transverse momentum, such as rapidity and multiplicity distributions, ratios 
of particles and it explains many characteristic features of hadron-hadron soft interactions.  
But for calculating pseudorapidity distribution from Eq.~(\ref{Eq:Eq2}) or (\ref{Eq:Eq4}) we 
must know mean transverse momentum of particles. 

\begin{figure}[h]
\label{Fig:Fig1}
\begin{center}
\resizebox{0.45\textwidth}{!}{%
\includegraphics{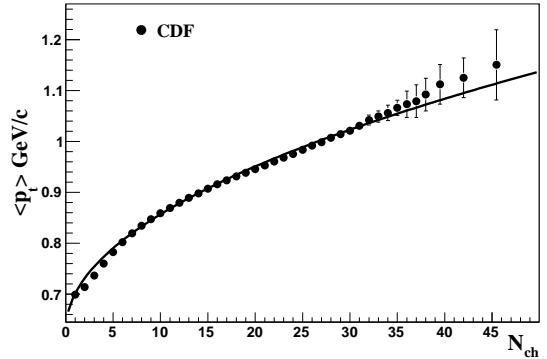}}
\caption{Comparison of phenomenologically expected dependence of mean transfers momentum of 
secondary particles on charged particles multiplicity with experimental data from 
CDF \cite{ref16}.} 
\end{center}
\end{figure}

Though the model does not give predictions on $p_t-$ –dependence, some conclusions can be 
obtained from its basics (or from Gribov's Reggeon calculus in general). As it is already 
mentioned, the multi-particle production is described in terms of cut-pomeron diagrams. 
On the other hand, using Eq.~(\ref{Eq:Eq1}) it is easy to see that the mean number of 
cut-pomerons increases with energy $\sim s^{\Delta}$. Average transverse momenta of 
produced particles increase with the number of chains: $p_t \sim \sqrt{N}$.  So, we expect 
at LHC energies to have mean transverse momentum of produced particles  higher then the 
one measured at ISR energies. Analogously, the selection of events with high multiplicity 
results in an increase of cut-pomerons number and to an increase of  $<p_t> \sim \sqrt{N}$. 
This dependence is observed experimentally by UA1 and CDF collaborations (\cite{ref15} and 
\cite{ref16}). Only in cases of \cite{ref16} the values of measurements are published and 
in Fig.~\ref{Fig:Fig1} we compare our expected dependence with these data. The fit result is:
\begin{equation}
p_t = (0.62973 \pm  0.00084) + (0.0717848 \pm 0.00024) \sqrt{N_{ch}}.
\end{equation}

In order to convert rapidity to pseudorapidity distributions we used experimentally 
measured $<p_t>$ for different types of particles  (here we assume that the sample 
of secondary charged particles consists from $\pi^{+}$, $K^+$ and protons, and their 
anti-particles). In Ref. \cite{ref8} we have parameterized $<p_t>$ as a function of 
$\sqrt{s}$ for charged pions, kaons and anti-protons, and found a successful fit to 
the data from ISR to Tevatron energies with the second-order polynomial function of 
$\ln(s)$. Fit result is \cite{ref8}:
\begin{eqnarray}
\label{Eq:Eq6}
<p_t^{\pi}> = 0.34 -0.002 \ln s + 0.00035 \ln ^{2} s,\nonumber\\
<p_t^{K}> = 0.55 -0.031 \ln s + 0.001  \ln^{2} s,\\
<p_t^{p}> = 0.65 -0.045 \ln s + 0.0036   \ln^{2} s.\nonumber
\end{eqnarray}
In the following for each energy and for each particle type we calculate $p_t$ using (\ref{Eq:Eq6}). 

\begin{figure}[h!]
\begin{center}
\resizebox{0.45\textwidth}{!}{\includegraphics{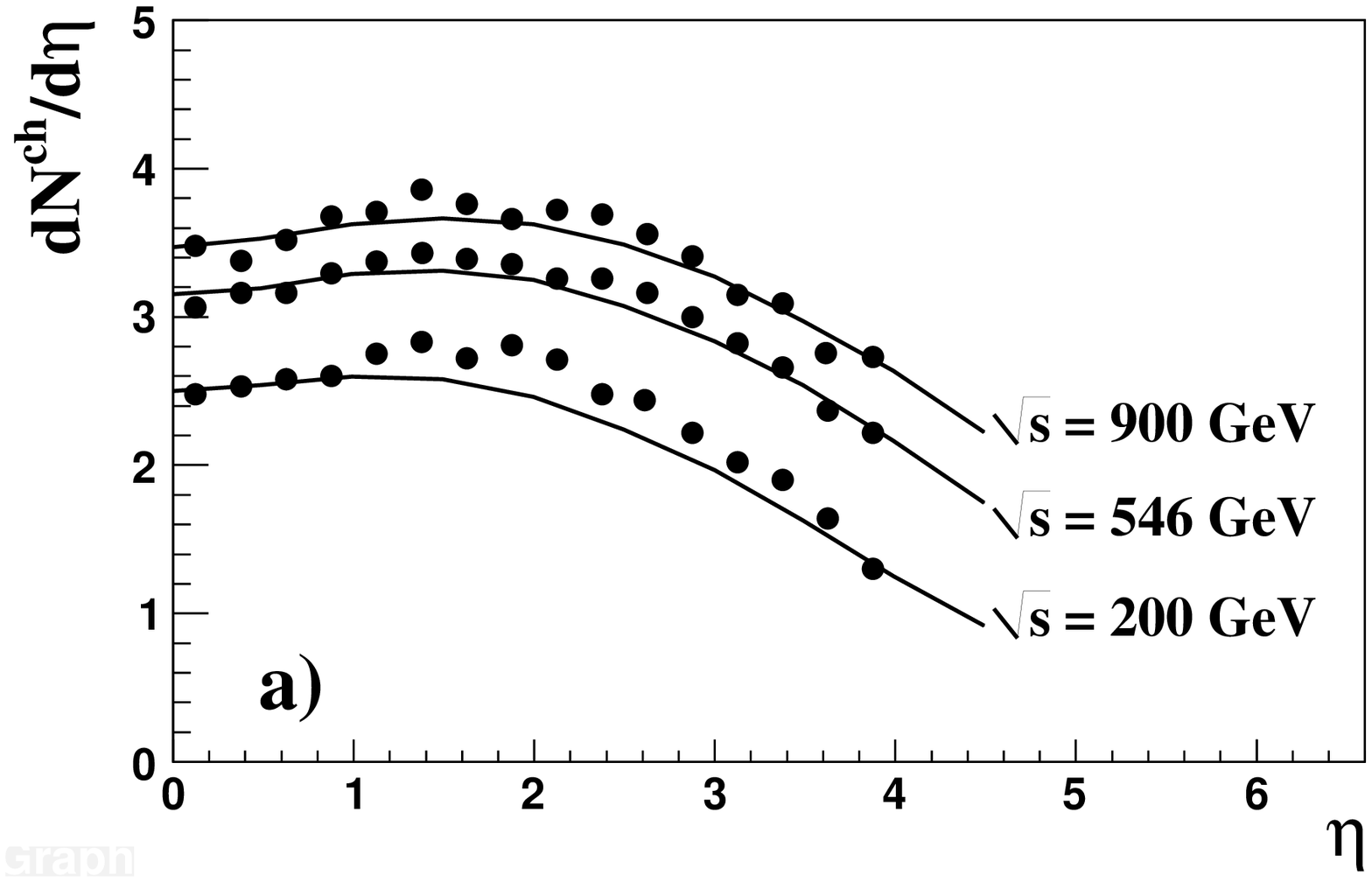}}
\resizebox{0.45\textwidth}{!}{\includegraphics{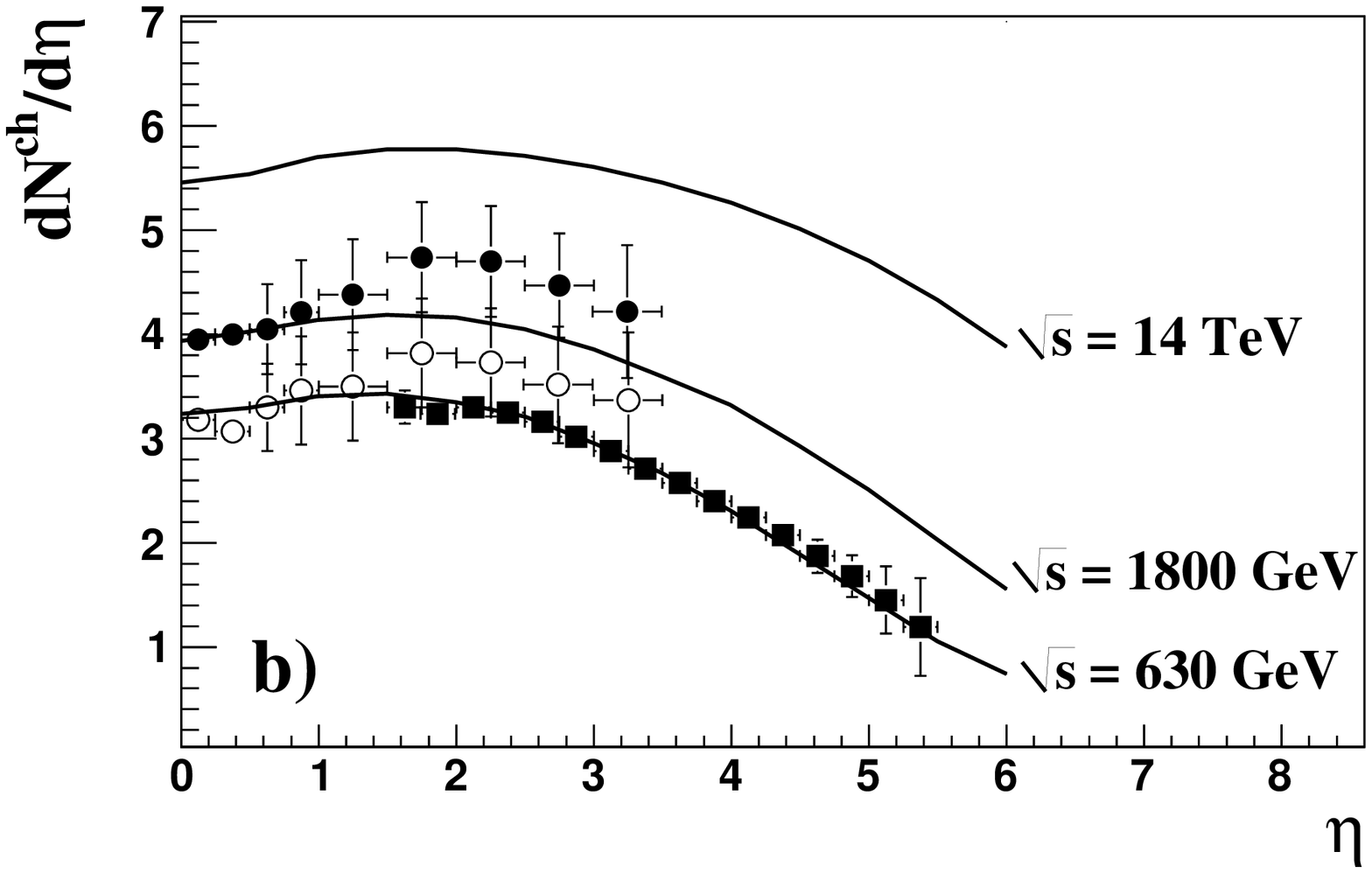}}
\caption{
Comparison of model's prediction with data on charged particles pseudorapidity distribution 
in NSD events and prediction for LHC: a) description of UA5 data \cite{ref17}, b) description 
of CDF and P238 data (circles and squares, respectively) \cite{ref18}.}
\label{Fig:Fig2}
\end{center}
\end{figure}

In Fig.~\ref{Fig:Fig2} we give description of S$p\bar{p}$S and Tevatron data on 
charged particles pseudorapidity distributions in $p\bar{p}$ non-single diffractive 
(NSD) events and give a prediction for $\sqrt{s}$ =14 TeV. 

\begin{figure}[h]
\begin{center}
\resizebox{0.45\textwidth}{!}{\includegraphics{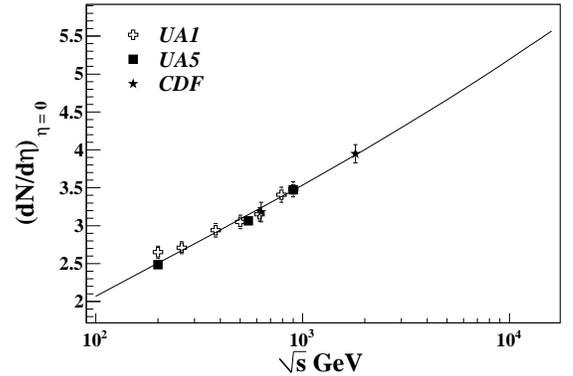}}
\caption{
Energy dependence of charged particles pseudorapidity density at $\eta =0$ for NSD events. 
Data are taken from \cite{ref17,ref18,ref19}.}
\label{Fig:Fig3}
\end{center}
\end{figure}

Energy dependence of charged particles pseudorapidity density for S$p\bar{p}$S NSD events in the 
central rapidity region is compared with data in Fig.~\ref{Fig:Fig3}. In the supercritical Pomeron 
($\Delta>0$) theory with account of ``non-enhanced'' diagrams (without interaction between Pomerons), 
which we consider now, inclusive cross-sections $d\sigma_c/dy$ at very high energies and at $y \simeq 0$ 
increase with energy as $(s/s_0)^{\Delta}$ (see the first article in Ref.~\cite{ref1} for more details). 
This means, in particular, that an energy dependence of inclusive spectra in the central rapidity region 
gives more reliable information on the value of $\Delta$, than an energy dependence of $\sigma_{tot}$, 
where pomeron cuts strongly modify energy dependence compared to the pole diagram.
An integral over rapidity density gives charged particles mean multiplicity $<N_{ch}>$. 
In Tables~\ref{Tab:Tab1}~and~\ref{Tab:Tab2} we list QGSM predictions on pseudorapidity density at mid-rapidity 
and mean number of charged particles for $pp$ interactions at various energies of LHC. Charged 
particles mean multiplicity is calculated for $|\eta|<1$ and $|\eta|<2.5$ pseudorapidity bins, 
corresponding to ALICE \cite{ref20} and ATLAS/CMS \cite{ref21} detectors central barrel acceptances, 
respectively, and for full phase-space. 
\begin{table*}
\caption{
Predictions for LHC on mean number of charged particles in two pseudorapidity bins and in full phase-space .
}
\label{Tab:Tab1}
\begin{tabular}[h!]{ccccccc}
\hline
$\sqrt{s}$ TeV &  $N^{NSD}(|\eta|<1)$ & $N^{Inel}(|\eta|<1)$ & 
$N^{NSD}(|\eta|<2.5)$ & $N^{Inel}(|\eta|<2.5)$ & $N^{NSD}$ & $N^{Inel}$\\

0.9 & 7.1 & 6.3 & 18 & 16  &  35.8 & 31.4\\
7   & 10.1& 8.9& 25.7& 22.6&  64.9 & 56.3\\
10  & 10.7& 9.4& 27.1& 23.8&  71.3 & 61.7\\
14  & 11.2& 9.8& 28.5& 25  &  77.4 & 67  \\
\hline
\end{tabular}
\end{table*}

\begin{table}
\caption{
Predictions for LHC on charged particles pseudorapidiy density at mid-rapidity.
}
\label{Tab:Tab2}
\begin{center}
\begin{tabular}[h!]{ccc}
\hline
$\sqrt{s}$ TeV & $(dN^{NSD}/d\eta)_{\eta=0}$ & $(dN^{Inel}/d\eta)_{\eta=0}$ \\
0.9 & 3.5 & 3. \\
7   & 4.9 & 4.2\\
10  & 5.2 & 4.5\\
14  & 5.5 & 4.7\\
\hline
\end{tabular}
\end{center}
\end{table}

\begin{figure}[h]
\begin{center}
\resizebox{0.45\textwidth}{!}{\includegraphics{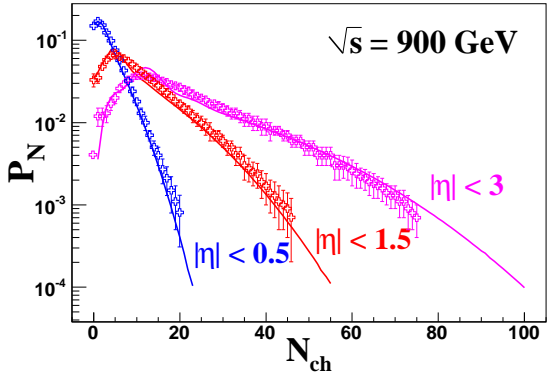}}
\caption{
Comparison of model predictions with UA5 data on charged particles 
multiplicity distribution in NSD events at $\sqrt{s}$ = 900 GeV.
}
\label{Fig:Fig4}
\end{center}
\end{figure}

\begin{figure}[h]
\begin{center}
\resizebox{0.45\textwidth}{!}{\includegraphics{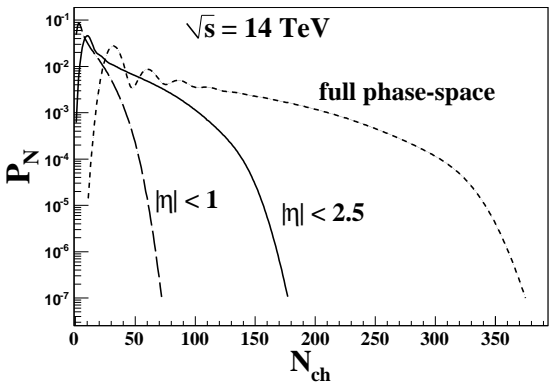}}
\caption{
Predictions for charged particles multiplicity distributions at LHC.}
\label{Fig:Fig5}
\end{center}
\end{figure}

The multiplicity distribution in the model is given according to Eq~(\ref{Eq:Eq2}) by a sum of $n$ 
cut-pomerons contributions and it is assumed Poisson-like form for each contributor. In this scheme, 
the KNO scaling \cite{ref22} is approximately valid up to ISR energies ($\sqrt{s} <$ 100 GeV, where 
the Poissonian distributions for a different number $n$ of cut-pomerons significantly overlap) and 
must be definitely violated at higher energies \cite{ref23}, as confirmed by measurements done 
at S$p\bar{p}$S and Tevatron.  In Fig.~\ref{Fig:Fig4} we give a description of UA5 data \cite{ref24} 
on charged particles multiplicity distribution at $\sqrt{s}$ = 900 GeV and for various pseudorapidity 
intervals. In Fig.~\ref{Fig:Fig5} predictions for $\sqrt{s}$ = 14 TeV for the full phase-space 
and for two preudorapidity bins 
(corresponding to ALICE and ATLAS/CMS detectors central barrel acceptances) are given. 

In this formulation, the model at very high energies (starting from $\sqrt{s} \sim $ 10 TeV) 
predicts $s^{\Delta}/\ln^2s$ asymptotic behavior for particles rapidity density at mid-rapidity 
and $s^{\Delta}/\ln s$ for mean number of secondary particles in the full phase-space. We stress 
that the contribution of enhanced diagrams (the interaction between Pomerons) that we do not take 
into account here are expected to be essential at these energies. With the account of these diagrams 
$(dN/dy)_{y=0} \sim ln^2s$ and $<N> \sim ln^3s$ asymptotic behaviors are expected. MC realization of 
the model with account of enhanced diagram is formulated in \cite{ref25}.

\begin{figure}[h]
\begin{center}
\resizebox{0.45\textwidth}{!}{\includegraphics{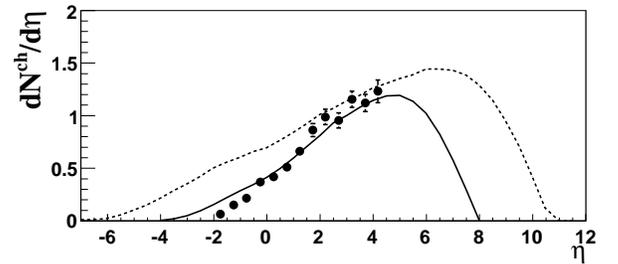}}
\caption{
Charged particles pseudorapidity distributions at $\sqrt{s}$ = 900 GeV (solid line) and 
14 TeV (dotted line) in singe diffractive dissociation events. The data points are from UA5.
The indicated errors are statistical and the systematical errorsareunknown.}
\label{Fig:Fig6}
\end{center}
\end{figure}

In Fig.~\ref{Fig:Fig6} we give a description of UA5 data \cite{ref26} on charged particles 
pseudorapidity distribution in single diffractive events at $\sqrt{s}$ = 900 GeV (dissociation 
of one of the colliding particles is considered) and give a prediction for $\sqrt{s}$=14 TeV. 
For both energies the spectra are calculated by integrating over all masses of the diffractive 
system (up to $M^2/s \leq$ 0.05). More comparisons with UA4 and UA5 data on SD events can be 
found in \cite{ref8}.

\section{Conclusion}
In this paper we compare QGSM predictions with S$p\bar{p}$S and Tevatron data on charged 
particles pseudorapidity and multiplicity distributions and give predictions for LHC. We 
stress that there are no free parameters in this analysis and hope that this approach will 
give a reliable predictions for particle production at LHC energies.

\section{Acknowledgements}
The work of A.B.K. was partially supported by the grants RFBR 0602-72041-MNTI, 0602-17012, 
0802-00677a and Nsh-4961.2008.2.

\end{document}